\documentclass[10pt]{iopart}
\usepackage[utf8]{inputenc}
\usepackage{amsfonts}
\usepackage{amssymb}
\usepackage{graphicx}
\usepackage{subfigure}
\usepackage{xspace}
\usepackage[pdftex,unicode=true,bookmarks=false,breaklinks=false,pdfborder={0 0 1},colorlinks=true]{hyperref}
\hypersetup{linkcolor=blue,citecolor=blue,urlcolor=blue}

\usepackage{setspace}

\newcommand{\celsius}{$~^{\circ}$C\xspace}

\begin{document}

\title{Surface preparation and patterning by nano imprint lithography for the selective area growth of GaAs nanowires on Si(111)}

\author{Hanno Küpers$^1$, Abbes Tahraoui$^1$, Ryan B. Lewis$^1$, Sander Rauwerdink$^1$, Mathias Matalla$^2$, Olaf Krüger$^2$, Faebian Bastiman$^1$, Henning Riechert$^1$, and Lutz Geelhaar$^1$}

\address{$^1$ Paul-Drude-Institut für Festkörperelektronik, Hausvogteiplatz 5--7, 10117 Berlin, Germany}

\address{$^2$ Ferdinand-Braun-Institut, Leibniz-Institut für Höchstfrequenztechnik, Gustav-Kirchhoff-Strasse 4, 12489 Berlin, Germany}

\begin{abstract}
	
The selective area growth of Ga-assisted GaAs nanowires (NWs) with a high vertical yield on Si(111) substrates is still challenging. Here, we explore different surface preparations and their impact on NW growth by molecular beam epitaxy. We show that boiling the substrate in ultrapure water leads to a significant improvement in the vertical yield of NWs (realizing 80\%) grown on substrates patterned by electron-beam lithography (EBL). Tentatively, we attribute this improvement to a reduction in atomic roughness of the substrate in the mask opening. On this basis, we transfer our growth results to substrates processed by a technique that enables the efficient patterning of large arrays, nano imprint lithography (NIL). In order to obtain hole sizes below 50~nm, we combine the conventional NIL process with an indirect pattern transfer (NIL-IPT) technique. Thereby, we achieve smaller hole sizes than previously reported for conventional NIL and growth results that are comparable to those achieved on EBL patterned substrates.

\end{abstract}

\ioptwocol
\section{Introduction}

In recent years, numerous electronic and optoelectronic devices based on semiconductor nanowires (NWs) have been demonstrated, including LEDs, lasers, and photovoltaic cells \cite{Dasgupta2014}. Fairly independently of the material, these structures can be grown directly on Si substrates, allowing direct bandgap III-V devices to be integrated with Si technology. For many applications, controlling the position of the NWs on the chip is essential. One prominent approach is the selective area growth (SAG) in the holes of a patterned mask, which is defined in thermal silicon oxide layers using advanced lithography methods. Due to the low sticking on the oxide surface, growth is restricted to the nano-holes. In the case of Ga-assisted growth of GaAs NWs by molecular beam epitaxy (MBE), this approach has been of great interest in recent years \cite{Bauer2010,Plissard2010,Gibson2013b,Munshi2014,Heiss2014}.

Despite much progress, realising a high vertical yield, i.e. ratio of vertical NWs to holes in the mask, remains challenging. Often, NWs form at the desired position but do not elongate perpendicular to the substrate or even crystallites form instead of NWs. Vertical yield values vary significantly among different studies \cite{Bauer2010,Plissard2011} because the yield depends not only on growth parameters \cite{Plissard2011,Munshi2014} but also critically on mask processing conditions \cite{Plissard2010,Gibson2013b}. 

Additionally, in core-shell NW devices \cite{Dimakis2014} it is desirable that the hole size is smaller than the NW diameter in order to minimize leakage currents between the substrate and the doped shells. Previously, two advanced lithography approaches have been used for the selective area growth of NWs: Electron beam lithography (EBL) \cite{Hersee2006,Motohisa2004180,Bauer2010,Plissard2010,Gibson2013b,Zhang2014d,Heiss2014} and nano imprint lithography (NIL) \cite{Pierret2010,Munshi2014,Hertenberger2012}. In principle, NIL is faster for large pattern sizes because EBL is a sequential process. However, so far the NIL approach could not realize feature sizes that are comparable to what was achieved with EBL (40~nm). Theoretically, the resolution limit of NIL depends mainly on the minimum feature size on the stamp, which can be fabricated by electron beam lithography. However, in practice, the precise pattern transfer into the mask layer with a high fidelity depends first on the thickness and the uniformity of the residual layer underneath the imprint pattern, and second on the optimization of the plasma etching parameters for each process step.

 In this study, we explore different surface preparation treatments and the impact of this processing step on the vertical yield. We show that boiling the wafer in ultrapure water can increase the vertical yield from 5\% to 65\% on EBL patterned substrates, indicating that the surface preparation is of crucial importance for the nucleation of GaAs NWs in the vapour-liquid-solid (VLS) mode. Based on this surface preparation, we can achieve a vertical yield of 80\% on EBL patterned substrates by further optimizing the growth parameters. In order to transfer this result to large arrays patterned by NIL we establish a process to reach hole sizes that are comparable to what we achieve with EBL. Our approach combines the conventional ultraviolet-NIL (UV-NIL) technique with an inverse pattern transfer process (NIL-IPT) \cite{NIL2011}. This novel combination enables the realization of holes with diameters below 50~nm, significantly smaller than values reported for the conventional NIL process with direct pattern transfer.

\section{EBL processing and growth experiments}

For the first part of this study, substrates were patterned by EBL. First, 100~nm of positive EBL resist was spin-coated on 2" and 3" Si(111) wafers covered with a 15--20~nm thick thermal silicon dioxide (SiO$_2$) layer. Then, the pattern was written in an EBL system. The EBL pattern comprises fields with hexagonal arrays of holes with pitches ranging from 0.1 to 10~$\mu$m and minimum hole diameters of 40--50~nm. Subsequently, the resist was developed and the oxide mask was etched by reactive ion etching using CHF$_3$. Finally, the wafers were cut into square pieces with an edge length of 10~mm and then cleaned by organic solvents, oxygen plasma and UV ozone. With this procedure we achieve 9 (29) highly comparable substrate pieces per 2" (3") wafer. Immediately before loading into the MBE system, the surface of the substrate was prepared by a wet chemical treatment as described in detail in the next section. 

The MBE system comprises effusion cells for Ga and a valved cracker source for As$_2$. The substrate temperature was measured by a pyrometer, calibrated to the oxide desorption temperature of GaAs(100). Prior to growth, substrates were annealed in the growth chamber at around 680\celsius for 10 minutes, after which the temperature was lowered to the growth temperature of 630\celsius . Ga was pre-deposited at a flux of 0.5~ML/s for 90~s. Subsequently, NW growth was initiated by supplying Ga and As$_2$ simultaneously at a V/III ratio of 2.4. The growth time was 15--30~min, after which all sources were closed and the substrate was ramped to 100\celsius . A more detailed description of the growth and related calibration routines can be found in our previous publication \cite{Bastiman2016}.

\begin{figure}
		\centering
		\includegraphics[width=\columnwidth]{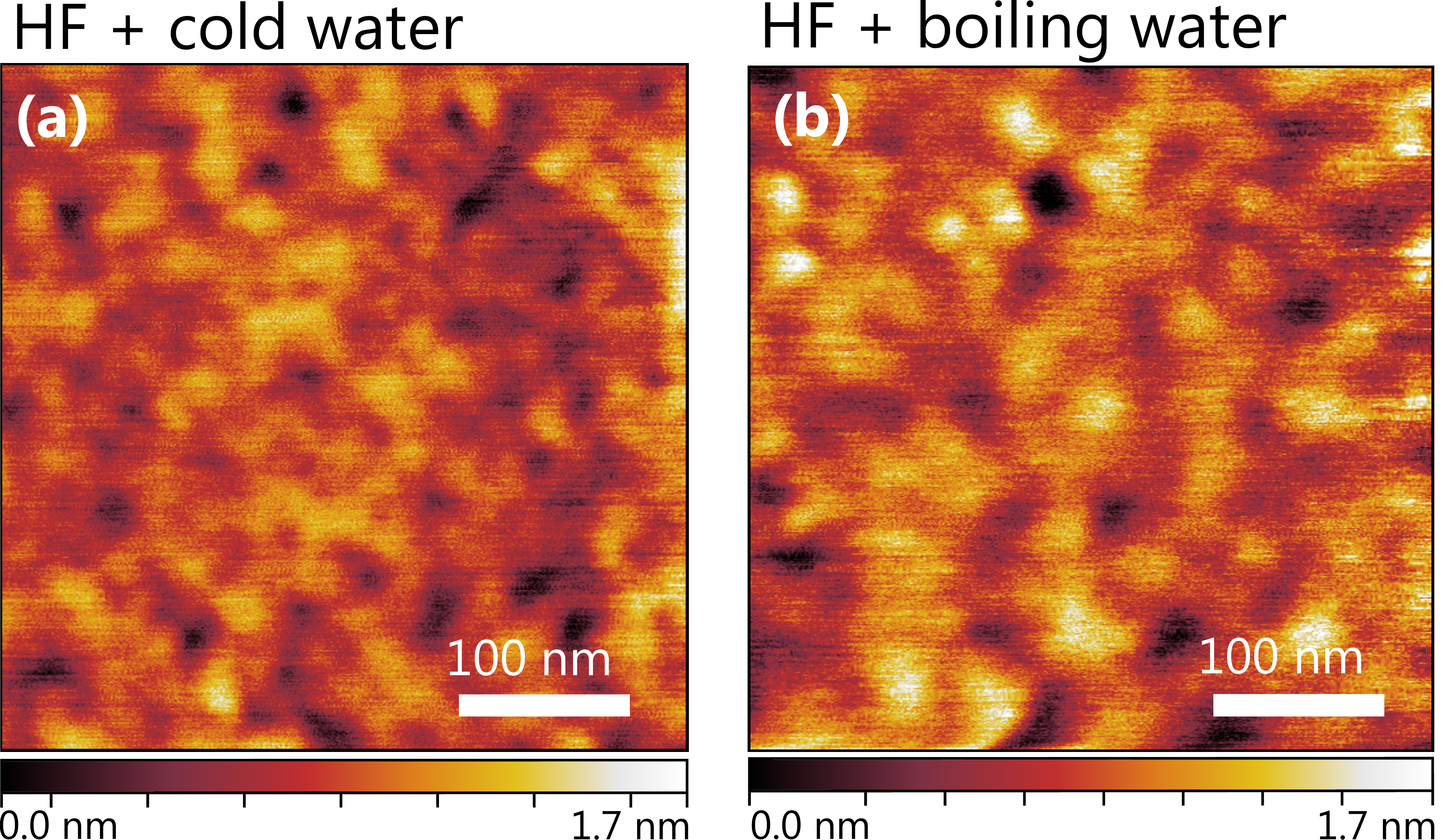}
		\caption{AFM micrographs of Si(111) surfaces treated with (a) HF and cold water and (b) HF and boiling water following the rinse in cold water. In both cases the measurement was carried out in large marker areas of a patterned substrate where the thermal oxide had been removed before the wet treatment.}
		\label{fig:WaterEtchAFM}
	\end{figure}
	
\begin{figure*}
		\centering
		\includegraphics[width=0.95\textwidth]{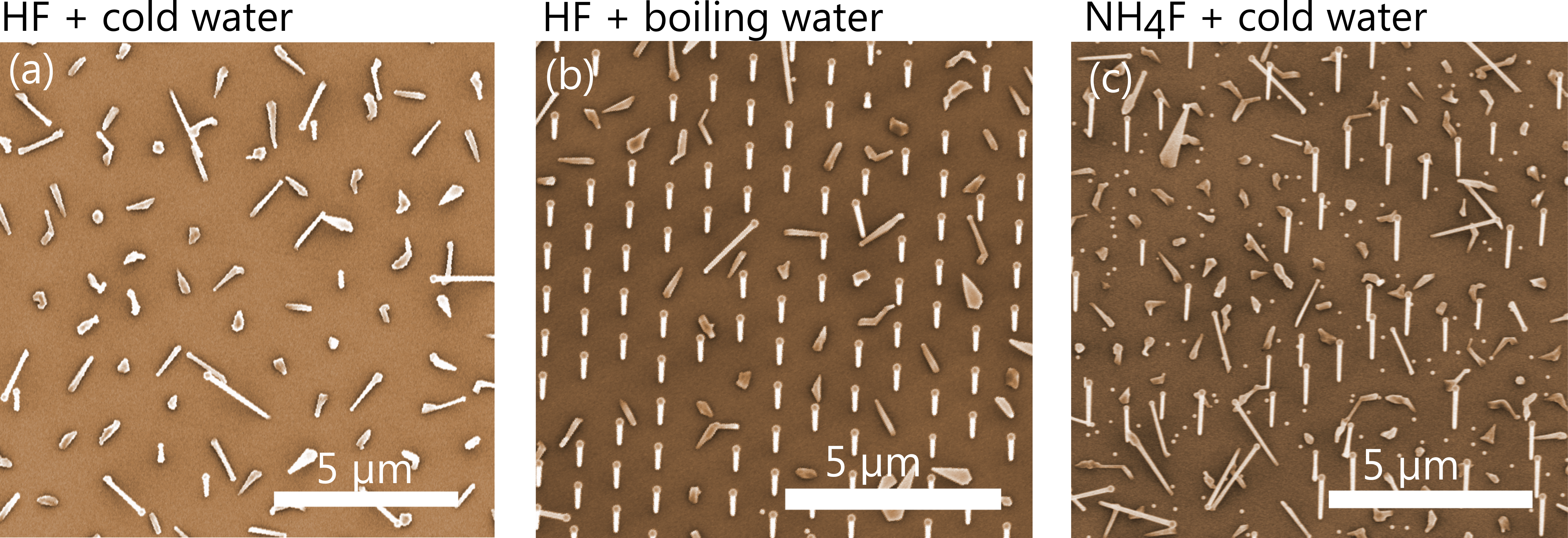}
		\caption{SEM micrographs comparing NW samples grown on substrates patterned by EBL and exposed to different surface treatments prior to growth: 
		(a) HF and cold water, (b) HF and cold and boiling water, and (c) NH$_4$F and cold water. The vertical yield increases from below 5\% to 65\% with the addition of the boiling water treatment. Also for the sample treated with NH$_4$F the vertical yield increases to 25\%. The viewing angle for all micrographs is 15$^\circ$ from normal.}
		\label{fig:WaterEtchSEM}
\end{figure*}

\section{Surface preparation}

One reason for the low reproducibility of the vertical yield of GaAs NWs in selective area growth by MBE is the limited understanding of the initial nucleation of NWs at the substrate-droplet interface. A way to change this interface is the exploration of different surface treatments of the Si(111) surface prior to growth. In general, fluoride acid solutions are employed to remove the native silicon oxide in the mask openings. Aqueous solutions of HF have been reported to produce atomically rough Si(111) surfaces~\cite{Higashi1990,Higashi1991}. The surface is oxide free but small clusters are present with di- and trihydrides saturating the dangling bonds of the Si atoms at the edges. These edges are selectively etched in etching solutions with a higher pH value, as for example ammonium fluoride (NH$_4$F), leading to an atomically flat surface \cite{Higashi1991}. A similar effect is achieved by boiling the sample in oxygen-free water for up to 10~min where OH$^+$ ions attack the Si backbonds \cite{Watanabe1991a}: 
It was reported that this treatment leads to a Si(111) surface which is completely terminated by mono-hydrides \cite{Watanabe1991}, and the smoothness of the surface on an atomic scale was confirmed by scanning tunneling microscopy \cite{Pietsch1992}.

Fig.~\ref{fig:WaterEtchAFM} presents the surface topography of etched marker areas as measured by atomic force microscopy (AFM) on patterned substrates after etching in 1\% HF solution for 60~s and rinsing with (a) cold (20\celsius) water and (b) cold and subsequently boiling (100\celsius) water. The root-mean-square roughness values are 0.19~nm and 0.17~nm, respectively. We cannot assume this difference to be significant due to the resolution limit of the setup. Even though we cannot access the atomic roughness by AFM measurements, the sample with the boiling water rinse shows a larger feature size (average equivalent square size is 31.2~nm in (a) and 38.0~nm in (b) as calculated by a segmentation grain analysis using \textit{gwyddion}). These larger islands are consistent with a smoother surface for the boiled sample.
	
In order to explore the impact of such surface treatments on NW growth, different treatments were carried out before loading the samples into the MBE system. Fig.~\ref{fig:WaterEtchSEM} shows scanning electron microscopy (SEM) images of samples after growth for substrates pre-treated with: (a) HF (1\%) for 60~s with 3~min cold (20\celsius) water rinse, (b) HF (1\%) for 60~s with 3~min cold (20\celsius) water rinse and 10~min hot (100\celsius) water rinse, and (c) NH$_4$F (40\%) for 120~s with 3~min cold (20\celsius) water rinse. The growth conditions were the same for all samples as described above. In Fig.~\ref{fig:WaterEtchSEM}~(a), the vertical yield is below 5\% with most holes occupied by tilted NWs or crystallites. However, the growth is restricted to the holes and the oxide surface seems to be free of residues. Fig.~\ref{fig:WaterEtchSEM}~(b) shows that adding a boiling water rinse in addition to the cold water rinse leads to a drastic increase in vertical yield to 65\%. Fig.~\ref{fig:WaterEtchSEM} (c) shows another sample which was grown on a wafer etched in NH$_4$F instead of the HF dip (no boiling water). This sample also exhibits an increase in vertical yield to 25\%. However, many droplets are present on the oxide surface indicating the presence of residues. Furthermore, the NWs of this sample have different lengths. Results from earlier experiments suggest that incompletely etched holes lead to the inhomogeneous length distribution. Here it may result from the low etching rate of the solution. The NH$_4$F etching leads to a smoother surface but was reported to leave insoluble salt residues on the surface \cite{Yang1994}. These residues may be the reason for the accumulation of material on the oxide surface as seen in Fig.~\ref{fig:WaterEtchSEM} (c).

\begin{figure}
		\centering
		\includegraphics[width=\columnwidth]{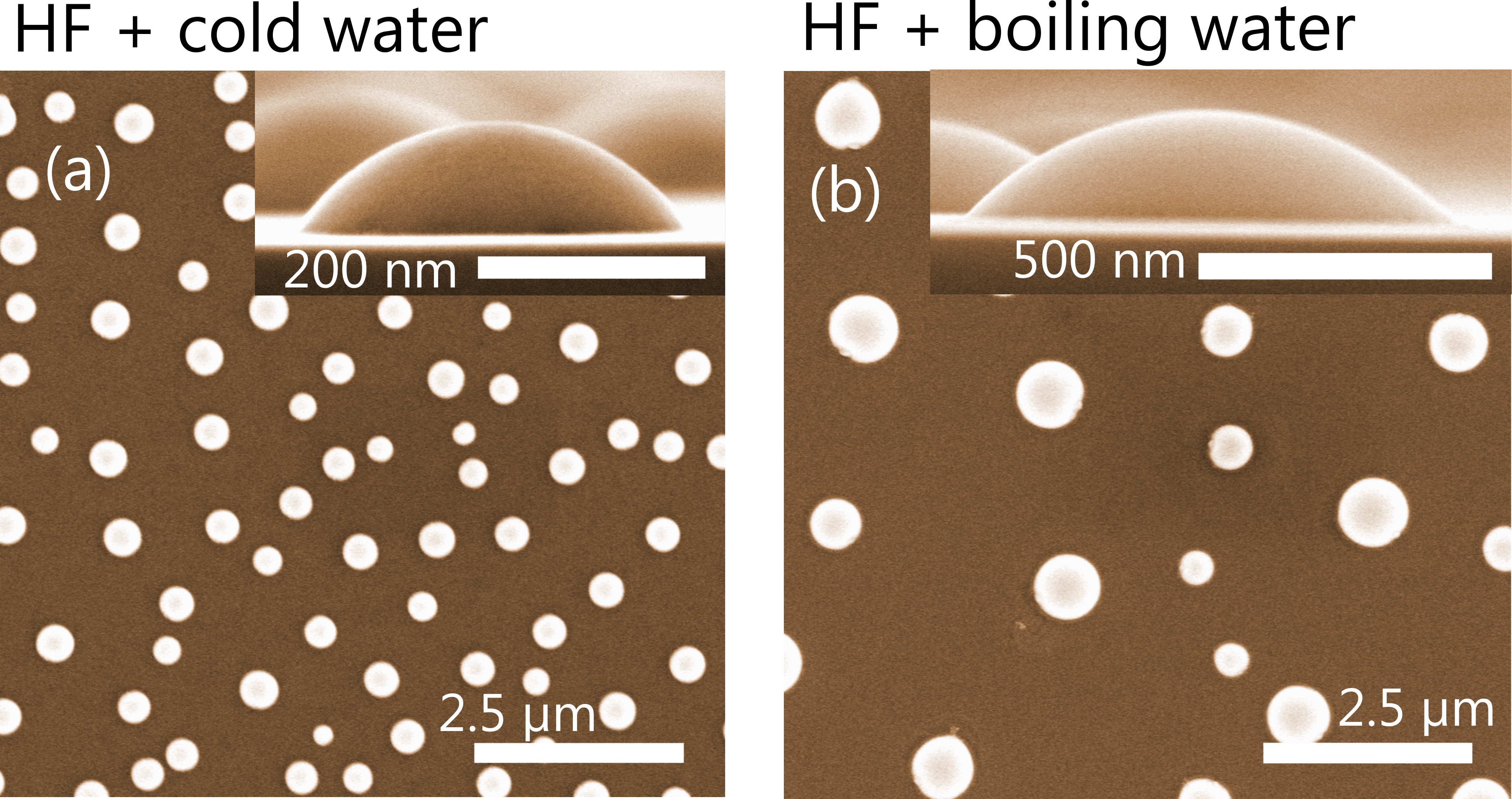}
		\caption{SEM top-view micrographs of Ga droplets deposited on unpatterned Si(111) substrates with different surface treatments: 
		(a) HF and cold water and (b) HF, cold water and boiling water.
Insets:	SEM micrographs in side-view, showing a contact angle of 50$^\circ$ and 45$^\circ$. }
		\label{fig:WaterEtchDroplet}
	\end{figure}

Previously, it was reported that the contact angle of droplets on the substrate surface can have a significant impact on the nucleation of NWs \cite{Matteini2015a}. In order to check if the here presented surface treatment changes the contact angle we deposited Ga droplets on unpatterned Si(111) substrates in a similar fashion as has been done in the mentioned study. Even though we assume that the hole in the oxide mask has a significant impact on the shape of the Ga droplet, here, we are interested in the surface properties of the Si substrate. This effect will be similar on a bare substrate and in an etched hole and therefore we can use unpatterned substrates for this experiment. Fig.~\ref{fig:WaterEtchDroplet} shows SEM top-view micrographs for samples with different surface treatments: (a) 1\% HF for 60~s and rinsing in cold water and (b) 1\% HF for 60~s and rinsing subsequently in cold and boiling water. The mean droplet diameter increases from 390 $\pm$ 50~nm to 620 $\pm$ 210~nm using boiling water and the density decreases from 0.92~$\mu m^{-2}$ to 0.21~$\mu m^{-2}$. The larger separation and size of the droplets indicate a longer surface diffusion length of Ga atoms on the Si surface for the substrate rinsed in boiling water, which is consistent with a smoother surface due to the hot water treatment. The insets of Fig.~\ref{fig:WaterEtchDroplet} show side-view micrographs of Ga droplets after the deposition. The contact angle is similar for the two samples (approximately 50$^{\circ}$ and 45$^{\circ}$). These values are in agreement with the reported values for an oxide free surface \cite{Matteini2015a}. Thus, the observed increase in vertical yield does not correlate with a significant change in contact angle. Consequently, NW nucleation cannot be understood by only investigating the contact angle and the underlying surface energies.

\begin{figure}
		\centering
		\includegraphics[width=0.65\columnwidth]{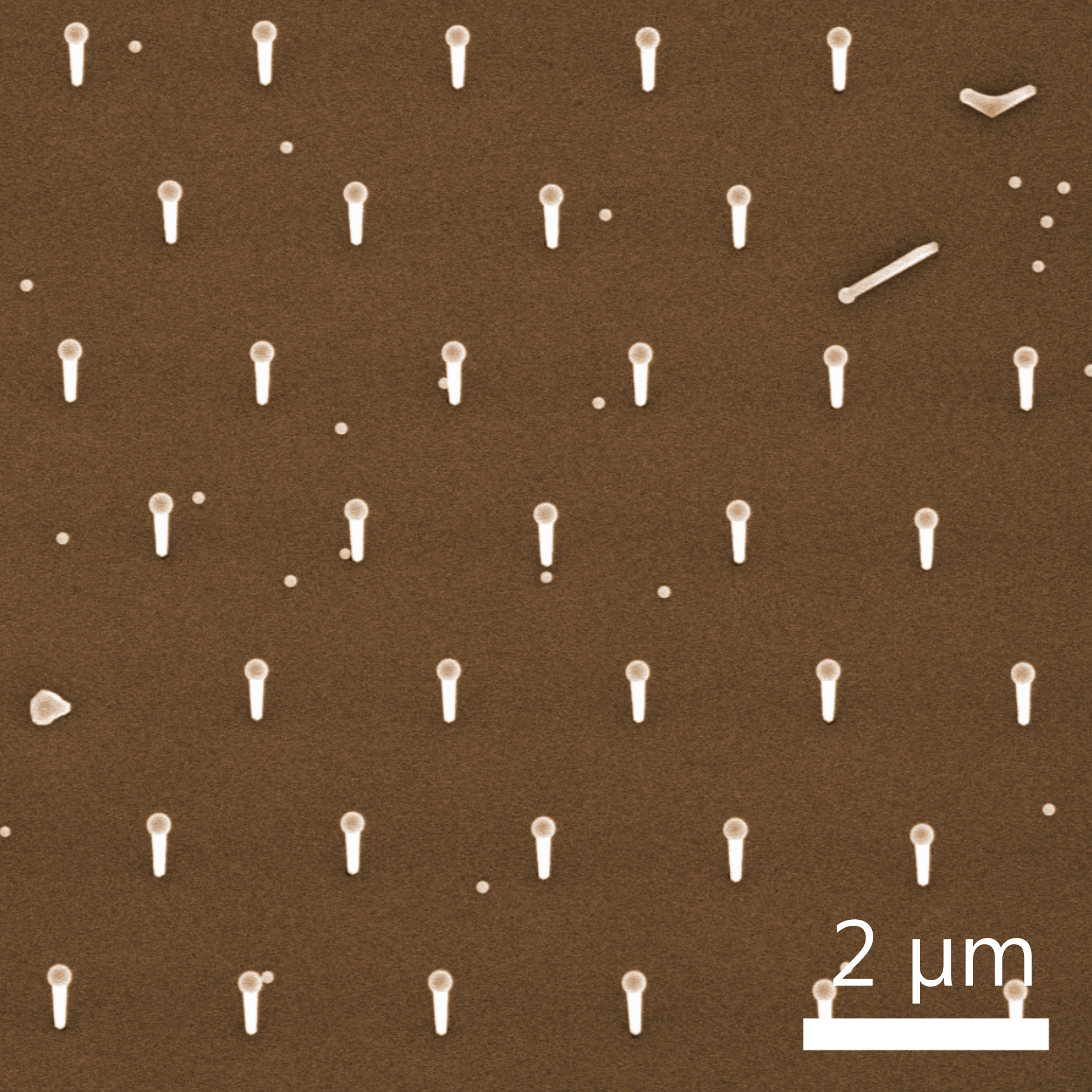}
		\caption{Micrograph of a NW sample grown on a substrate pre-patterned by EBL with optimized surface treatment and growth conditions. An overall vertical yield of 80\% was achieved.}
		\label{fig:optEBL}
	\end{figure}	

After further optimization of the growth parameters we could achieve a vertical yield of 80\% as seen in Fig.~\ref{fig:optEBL}. Our results underline the importance of the surface preparation for NW nucleation. Furthermore, our results are in agreement with the hypothesis that a smoother substrate surface leads to an improved vertical yield. We suppose that an atomically rough substrate surface may lead to a high density of initial nuclei at the droplet substrate interface leading to a rapid crystallization of the liquid Ga droplet into a GaAs crystallite.

\section{NIL with inverse pattern transfer}

\begin{figure*}
		\centering
		\includegraphics[width=0.85\textwidth]{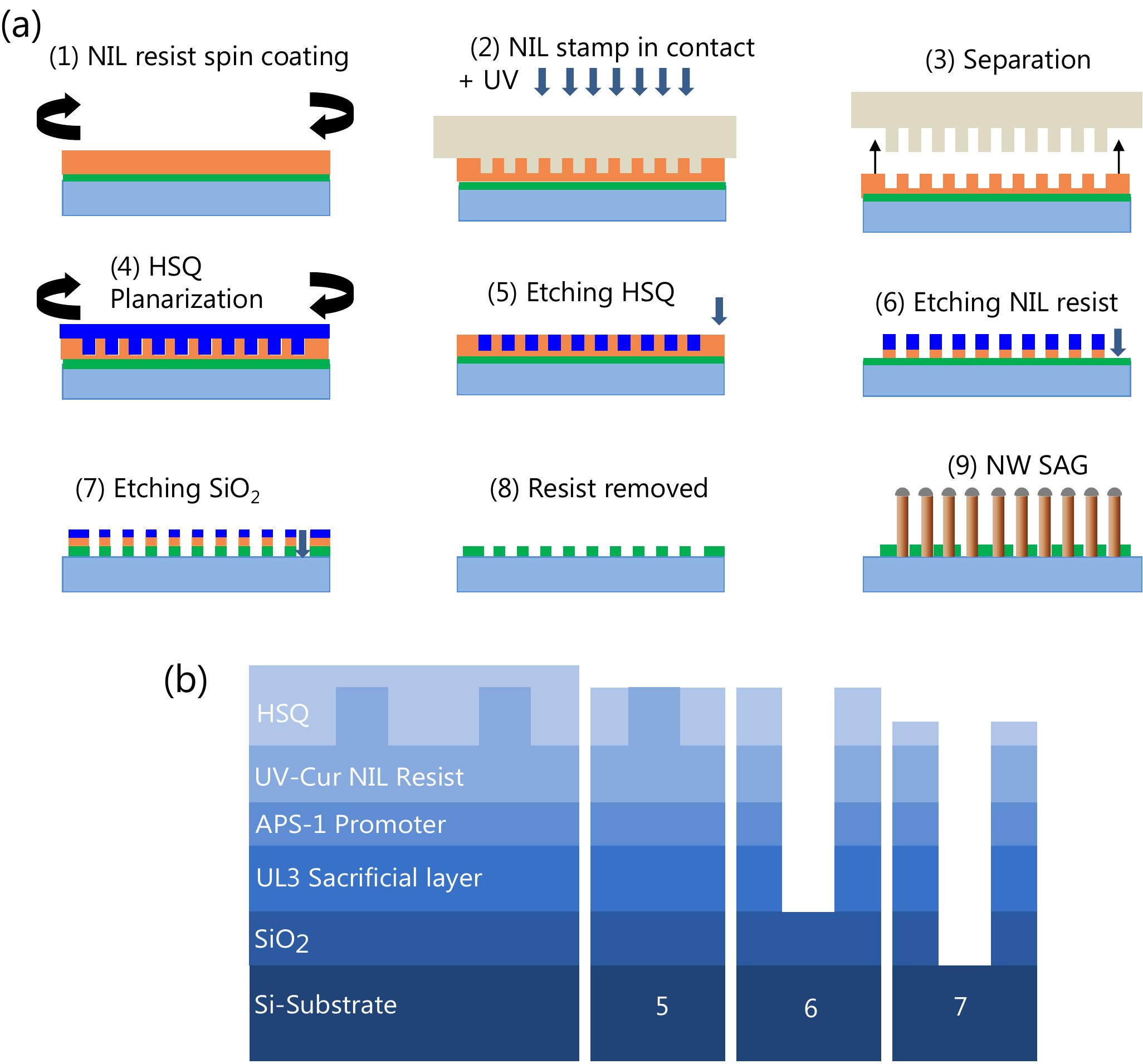}
		\caption{(a) Diagram of the workflow in NIL-IPT: (1) Deposition of processing layers by spin-coating, (2,3) formation of nanopillars in the NIL resist, (4) planarization of pattern by deposition of HSQ layer, (5) back-etching of HSQ layer, (6) back-etching of NIL resist to form etching mask, (7) etching of oxide mask, (8) lift-off process to remove processing layers and final cleaning, (9) substrate ready for NW growth.
		(b) Diagram depicting the different layers of resist material and the structure after the three etching steps 5--7.}
		\label{fig:NILworkflow}
\end{figure*}	

For the realization of large NW arrays, it is desirable to transfer the growth from EBL patterned substrates to substrates patterned by NIL-IPT. The complex process is depicted in Fig.~\ref{fig:NILworkflow} (a). First, the Si(111) wafer with 20 nm thermal silicon oxide film is cleaned with oxygen plasma for 10 min in order to remove organic residue and enhance the hydrophilicity of the surface. Then, a 180 nm thick sacrificial resist layer UL3 is spin-coated (step 1 in Fig.~\ref{fig:NILworkflow}), which is used for a lift-off process at the end of processing. Next, a thin film of adhesion promoter (mr-APS1) is spin coated and annealed at 150\celsius for 1 min. Subsequently, the imprint resist (\mbox{[mrUV-Cur21]}) is spin-coated and the sample is soft baked for 1 min at 80\celsius , in order to create a uniform layer and to remove solvent residues. 

After coating the wafer with all resist layers, the NIL stamp is used to transfer the pattern into the UV-NIL resist layer (2). The stamp is made out of UV-transparent quartz and consists of arrays of nanoholes. Such a stamp is more robust compared to a stamp containing nanosized lamellas to directly imprint holes in the resist. Therefore, its life time is increased significantly, which helps prevent the creation of defects during the release of the stamp from the imprint sample and improves the cleaning process. Before the contact step, the surface of the UV-NIL stamp is treated with oxygen plasma for 10 min and coated with an anti-sticking solution. During the imprint process, the quartz stamp is horizontally levelled to the substrate-resist system and pressed onto the film with low pressure at room temperature. A helium gas flow is applied during the contact phase to avoid trapping of air bubbles between the stamp and the resist \cite{Liang07}. While the stamp and the substrate are in contact, the resist is exposed to UV light through the stamp, which promotes crosslinking in the resist. After having formed nanopillars in the resist layer, the mask is released (3) and a 200~nm thick hydrogen silsequioxane layer (HSQ) is deposited to planarize the sample surface (4). The final structure of the resist layers is depicted in Fig.~\ref{fig:NILworkflow}~(b).

The inverse pattern transfer consists of three different etching steps using plasma reactive ion etching. First, the HSQ layer is back etched under CHF$_3$ gas down to the top of the nanopillars (5). In the second step, these pillars are selectively etched down to the SiO$_2$ surface under oxygen plasma at \mbox{-20\celsius} at a pressure of 0.8 Pa (6). In the third step, the pattern is transferred into the SiO$_2$ layer by CHF$_3$ plasma etching (7). During this etching step the top HSQ layer is also partly etched but the etching of the SiO$_2$ layer is not affected. As the last processing step, the resist layers are removed by a lift-off process using DI-water and the samples are cleaned by organic solvents, oxygen plasma and UV ozone (8). We want to mention that all resist materials used in the NIL-IPT process are purely organic provided by Microresit Technology. In combination with the lift-off process that is promoted by solution in DI-water, this feature assures the final surface cleanliness and the conservation of the underlying layer. Finally, these substrates are used for NW growth (9).

\begin{figure}
		\centering
		\includegraphics[width=0.65\columnwidth]{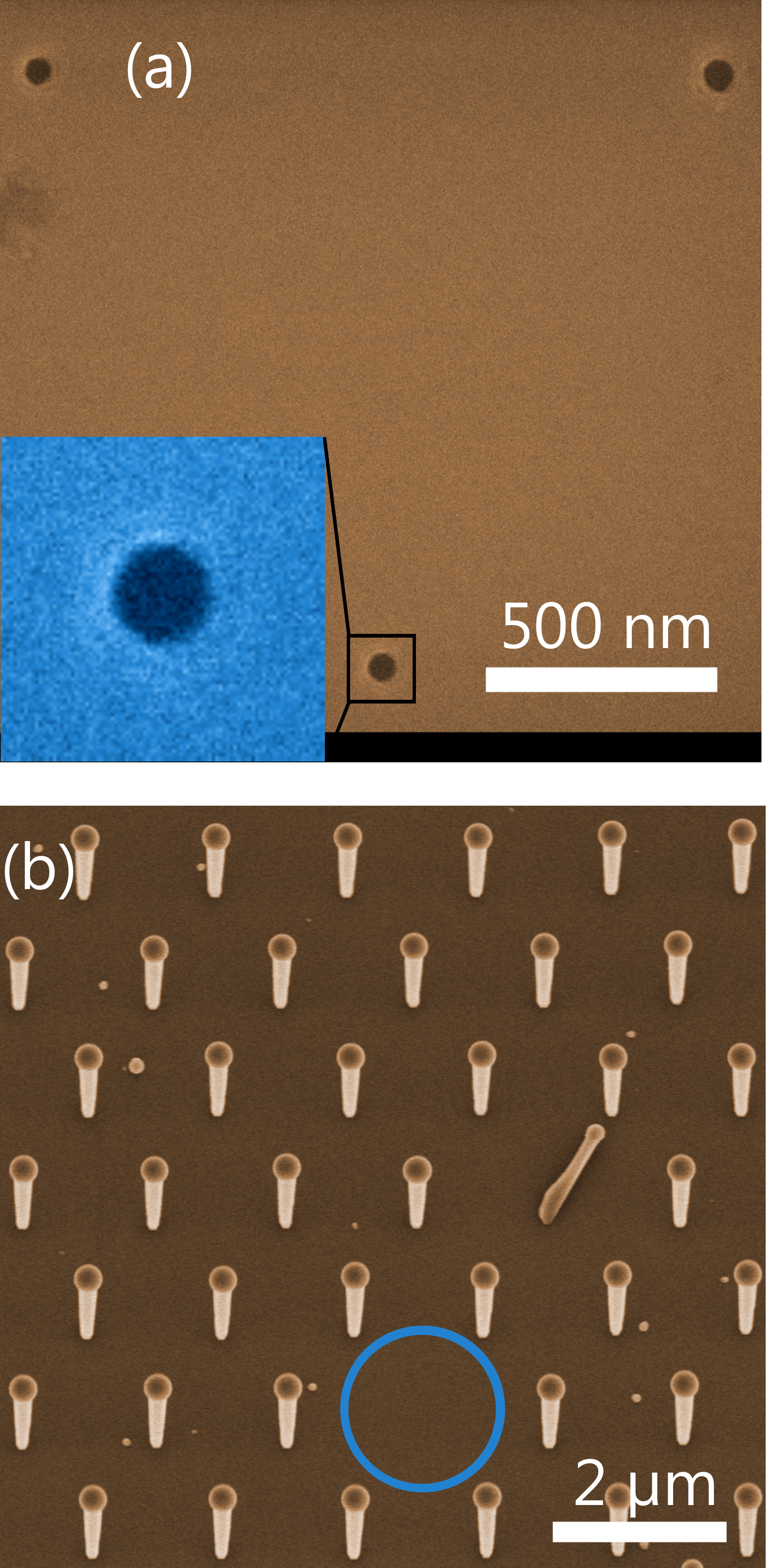}
		\caption{Micrographs of substrates patterned by NIL-IPT with inverse transfer (a) before and (b) after growth. The inset in (a) shows a larger view of one hole, emphasizing the small hole size of below 50~nm. The box has a size of 200 x 200 $nm^2$. The NW sample was tilted by 15$^{\circ}$ from normal.
		}
		\label{fig:LitComparison}
	\end{figure}	

Fig.~\ref{fig:LitComparison} (a) shows high resolution SEM images of the final pattern surface of substrates processed using the presented process. Hole diameters of below 50~nm could be realized with high fidelity. This result shows a clear improvement in the hole sizes of arrays used for NW growth compared to recent results using NIL with direct pattern transfer, showing hole diameters of 60~nm \cite{Hertenberger2012} and 100~nm \cite{Munshi2014}. In the direct approach holes are directly imprinted into the etching mask. During the etching of the thin residual layer that forms during imprint between stamp and substrate, the isotropic nature of the plasma etching leads to a widening of the hole size and inhomogeneities of the imprint pattern. This effect can be avoided by using the indirect pattern transfer, where the HSQ layer prevents hole widening during the second selective etching step. Even though our approach allows for smaller holes, we need to state that the complexity of the process makes it very sensitive to disturbances and it requires much time for process implementation and optimization.

Fig.~\ref{fig:LitComparison} (b) shows a SEM image of a NW array grown on a substrate patterned by the presented NIL-IPT process. The same growth conditions and surface treatment have been used that led to the optimized result for EBL patterned substrates as seen in Fig.~\ref{fig:optEBL}. The vertical yield of the NIL-IPT patterned sample is comparable (above 80\%) to what we obtained for EBL processed substrates, and to results on substrates patterned by NIL with direct transfer \cite{Munshi2014}, displaying the efficacy of both processes for selective area growth of NWs. However, for the array grown on the NIL-IPT processed substrate a hole in the pattern is missing (blue circle in Fig:~\ref{fig:LitComparison}). This defect of the pattern regularly appears on NIL-IPT samples. It is caused by trapping of air bubbles between the stamp and the resist during the contact step, which results in missing pillars during the pattern transfer. One way to overcome this issue might be to use the step-and-flash imprint lithography \mbox{(S-FIL)} approach \cite{Colburn99}, in which the exact amount of the dispensed droplets is controlled and optimized,  which helps to reduce the thickness of the residual resist layer significantly after the separation step, leading to an imprint layer with improved homogeneity.

\section{Conclusions} 

We improved the vertical yield of Ga-assisted GaAs NWs grown by MBE on pre-patterned substrates from 5\% to 65\% by following an improved substrate preparation procedure. The key process is rinsing in boiling water as the last step before loading the substrate into the MBE chamber. The origin for the improvement is not clear but we expect that it is related to the atomic scale roughness. These results will be important for understanding nucleation of VLS NWs and will help facilitate the reproducible and comparable selective area growth of VLS NWs.

Furthermore, using NIL with an inverse pattern transfer (NIL-IPT) we realized hole sizes smaller than those reported previously with NIL with direct pattern transfer, in particular below 50~nm. After optimization of the growth conditions we achieved vertical yield values of above 80\% for substrates patterned by EBL and NIL-IPT. Therefore, this study presents the basis for the growth of NW samples on large-scale substrates and cost-effective patterns with a high vertical yield.

\section{Acknowledgements}
This work was partially funded by Deutsche Forschungsgemeinschaft under grant Ge2224/2 and by the Alexander von Humboldt Foundation. We are grateful to Anne-Kathrin Bluhm for acquiring SEM images and to Michael Höricke, Carsten Stemmler and Arno Wirsig for technical support at the MBE system as well as to Bernd Drescher for support with the sample processing. We appreciate the critical reading of the manuscript by Alberto Hern\'{a}ndez-M\'{i}nguez.

\section*{References}

\bibliography{libSAGtech}

\end{document}